\documentclass[manuscript]{acmart}
\AtBeginDocument{%
  \providecommand\BibTeX{{%
    \normalfont B\kern-0.5em{\scshape i\kern-0.25em b}\kern-0.8em\TeX}}}

\setcopyright{none}

\begin{document}

\title{From 996 to 007: Challenges of Working from Home During the Epidemic in China}

\author{GAO Jie}
\email{jie_gao@mymail.sutd.edu.sg}
\affiliation{
  \institution{Singapore University of Technology and Design}
  \city{Singapore}
  \country{Singapore}
}

\author{Pin Sym FOONG}
\email{pinsym@gmail.com}
\affiliation{
  \institution{National University of Singapore}
  \city{Singapore}
  \country{Singapore}
}

\author{YANG Yifan}
\email{yangyifan46@126.com}
\affiliation{
  \institution{University of Nottingham Ningbo China}
  \city{Ningbo}
  \country{China}
}

\author{JIANG Weilin}
\email{21951221@zju.edu.cn}
\affiliation{
  \institution{Zhejiang University}
  \city{Hangzhou}
  \country{China}
}

\author{CHEN Yijie}
\email{ychen53@inside.artcenter.edu}
\affiliation{
  \institution{ArtCenter College of Design}
  \city{Pasadena}
  \country{USA}
}

\author{YING Xiayin}
\email{yingxiayin@outlook.com}
\affiliation{
  \institution{Zhejiang University}
  \city{Hangzhou}
  \country{China}
}

\author{Simon Perrault}
\email{simon_perrault@sutd.edu.sg }
\affiliation{
  \institution{Singapore University of Technology and Design}
  \city{Singapore}
  \country{Singapore}
}

\begin{abstract}
  During the COVID-19 epidemic in China, millions of workers in tech companies had to start working from home (WFH). The change was sudden, unexpected and companies were not ready for it. Additionally, it was also the first time that WFH was experienced on such a large scale. We used the opportunity to describe the effect of WFH at scale for a sustained period of time. As the lockdown was easing, we conducted semi-structured interviews with 12 participants from China working in tech companies. While at first, WFH was reported as a pleasant experience with advantages, e.g. flexible schedule, more time with family, over time, this evolved into a rather negative experience where workers start working all day, every day and feel a higher workload despite the actual workload being reduced. We discuss these results and how they could apply for other extreme circumstances and to help improve WFH in general.
\end{abstract}



\maketitle

\section{Introduction}
Remote work (or teleworking, telecommuting)~\cite{bailey2002review, crosbie2004work, nakrovsiene2019working, bloom2015does, siha2006telecommuting} has been practiced for many years all around the world.
In the main economies of the world, US and EU have respectively 3.6\%\footnote{\url{https://globalworkplaceanalytics.com/telecommuting-statistics}} and 5.1\%\footnote{\url{https://ec.europa.eu/eurostat/web/products-eurostat-news/-/DDN-20180620-1}} remote workers in the entire workforce in 2018.
In contrast, there are few data about China’s estimated proportion of remote workers and we can infer from the scarcity of data about the gap between China and other economies. This gap was likely narrowed during the COVID-19 epidemic\footnote{We use epidemic in this paper to signal the timing of the study, which covered events before 11th of March 2020, when the World Health Organization declared a global pandemic.} in China, forcing companies to develop WFH arrangements.
By the end of the Chinese New Year holiday (February 2020), there were an estimated 200 million people WFH\footnote{\url{http://www.Chinanews.com/business/2020/02-03/9077412.shtml}}. << Define 996 here >>

Previous research about remote work has focused on how workers experience it~\cite{martin2012telework, van2019no, dambrin2004does, gajendran2007good, baruch2001status}, for example, saving time and money, building work-life balance and so on. One limitation of previous studies was the lack of prevalent data of workplace flexibility arrangements~\cite{later2014executive}. To address the gap on remote work in China, Bloom et al.~\cite{bloom2015does} conducted a large-scale ($N=249$) empirical experiment in 2012 in China. The results indicated that employees' performance improved by 13\% during their experiment. 

However, we argue that these results were done for a target population whose work is easily quantifiable, which reduces challenges such as management regulations, and ultimately affect workers' attitude or willingness~\cite{ismail2019modelling} towards WFH. Furthermore, the positive outcome was not coherent with the low adoption rate of remote workers in China. Also, extreme contexts like the epidemic have been previously shown to make employees' experience extremely different from normal ones by impacting their mental health~\cite{brooks2020psychological, johal2009psychosocial} in areas such as feelings of isolation and loneliness and anxiety about social connections. Based on these studies, we believe it is important to document and describe the experience of working from home (WFH) in China. We wanted to know how employees cope with difficulties such as efficiency issues, and what organizational supports were available for employees. Specifically, our research questions are as follows:

\begin{itemize}
\item  \textbf{RQ1:} What were people’s WFH lives like during COVID-19 lockdown in China? 
\item \textbf{RQ2:} What kind of difficulties did Chinese people WFH encounter? 
\item \textbf{RQ3:} What were the main challenges during the epidemic and how to address them for better WFH practice in the future?
\end{itemize}

To answer these questions, we used the enforced lockdown during the epidemic in China as an opportunity to collect and analyze data. We conducted semi-structured interviews with 12 participants to study how the sudden switch from office work to WFH during the period from 24th of January to 30th of April 2020 affected Chinese people’s experiences of WFH. Our interviews show that after an initial positive image of WFH, numerous issues started appearing, e.g. inefficiency of online communication, extended working hours, increased interruptions, leading to overall lower work efficiency and increased levels of stress in addition to the feeling of being working 24 hours per day, 7 days per week (007). We contribute to improving the knowledge on WFH by identifying the challenges of people encountered during the epidemic and providing implications for design in the WFH area.

\section{Background and Related work}
\subsection{WFH's Definition and Reliance on Digital Technologies}
In this part, we reviewed studies in remote work to gain a better understanding of our research domain~\cite{bailey2002review, crosbie2004work, nakrovsiene2019working, bloom2015does, siha2006telecommuting, brocklehurst2001power, raghuram2019virtual, nilles1975telecommunications, nilles1997telework, aguilera2008business}. In the literature, WFH is one of the different forms of teleworking~\cite{siha2006telecommuting, raghuram2019virtual}, which describes a process where people use information and communication technologies~\cite{nilles1975telecommunications} to work at home rather than at centralized office location, or at any other dispersed spaces such as Starbucks or McDonald’s. Different terms used to describe this work seem to vary by frequency of this arrangement, ranging from regular teleworking~\cite{bailey2002review, nilles1997telework}, mobile workers~\cite{aguilera2008business} and occasional home workers ~\cite{bailey2002review}. The most cited definition of teleworking is from Nilles~\cite{nilles1997telework}. He specifies teleworking as \textit{the use of information technology to partially or totally replace work-related travel}. The definition of this form is broad and does not define the work location. In contrast, we propose to use the term WFH, which has a clear location where the work is conducted, to describe our research focus. 

Notably, although the forms of remote work are different, all of them describe a reliance on digital technologies~\cite{nilles1975telecommunications}. Digital technologies include digital hardware and software (e.g. computer, phone, internet, big data, cloud computing and so on) that support remote working operation. The effect of these technologies on WFH is emphasized from different perspectives~\cite{miele2020digital, raghuram2019virtual, ruilleryou, nilles1997telework}. Similar to Nilles, ~\cite{raghuram2019virtual} sees computer-mediated communication, as the basic requirement of dispersed teams’ communicating and interacting. ~\cite{miele2020digital} stated that digital technologies are employed to shape power and control practices within workplaces including remote and normal work.  ~\cite{ruilleryou} posits that the use of digital technologies provides permanent links between teleworkers and workplaces and improves the quality of life by allowing to redesign the working hours. This body of work emphasizes the important roles of digital technologies have played in WFH.

\subsection{Background and Context of WFH in China during the Epidemic}
According to BBC News\footnote{\url{https://www.bbc.com/news/world-52103747}}, over 100 countries and districts, including Europe, Japan, Thailand,  had implemented a full or partial lockdown to prevent the spread of COVID-19 by the end of March 2020. As a result, more than 250 million people were in lockdown in Europe as of 18 March\footnote{\url{https://en.wikipedia.org/wiki/2019\%E2\%80\%9320_coronavirus_epidemic\#Socioeconomics}}. Similarly, China also implemented an immediate and strictly enforced lockdown at the end of January and beginning of February 2020 which coincided with the largest public holiday period - Chinese New Year or the Spring Festival. The extended public holiday made the start of the lockdown somewhat easier, as organizations and workers typically extend this official holiday period by 1-2 weeks. Taking advantage of this long period, Chinese people working in the megacities such as Shanghai and Beijing typically journey to their hometowns across China and stay with their family members for the festive season. Then they return to their work cities again after this holiday. In this year, however, a large part of people could not go back to their work cities after CNY and were forced to extend their stay with their family members (mainly parents) during WFH. As a result, by the end of the holiday, around 200 million people were working remotely with their families.

Fortunately, the explosion of digital technology adoption in China during the past decade made it easier for the millions of workers who were forced to stay at home.  Prior to COVID-19, WFH was also used as a social isolation tool during SARS in 2003\footnote{\url{https://www.nytimes.com/2003/05/03/world/sars-epidemic-businesses-executives-singapore-chafe-sars-related-travel-bans.html}} and H1N1 in 2009-2010\footnote{\url{http://www.stats.gov.cn/tjsj/ndsj/2019/indexch.htm}}. Although there we could not find research which detailed the experience of WFH during those periods, we can infer from the basic network infrastructure data that the digital technologies and internet were not developed enough to support the operation of WFH in 2003 and 2009. For instance, according to the statistics released by China Internet Network Information Center (CNNIC), there were 4.9\%  netizens (68/1360 million) in total population in 2003\footnote{\url{http://www.cac.gov.cn/files/pdf/hlwtjbg/hlwlfzzkdctjbg012.pdf}}\footnote{\url{http://www.stats.gov.cn/tjsj/ndsj/2019/indexch.htm}} while around 65\% netizens (904/1400 million\footnote{\url{http://www.xinhuanet.com//2020-01/17/c_1125474664.htm}}) in 2020 in China~\cite{cnnic2014statistical}. Further, the same reports show that among these netizens 97.1\% had to use Desktop PCs to access the internet in 2003 while in contrast, internet access was available to 99.3\% on their smartphones by March 2020. Therefore, both the quantity and availability of digital technologies developed explosively in the past 17 years, which provides good conditions for the adoption WFH. 

\subsection{Challenges Brought by New Context}
Even before the epidemic, researchers have examined the pros and cons of WFH ~\cite{felstead2001working, siha2006telecommuting, fogarty2011half}. On the one hand, WFH is a good way to solve work-family conflicting needs~\cite{gajendran2007good}, save commuting time~\cite{nilles1988traffic, ismail2019modelling}, improve productivity~\cite{martin2012telework, bloom2015does}, increase time flexibility~\cite{ruilleryou} and so on. On the other hand, WFH also has raised many issues and challenges, such as management challenges~\cite{ruilleryou, dambrin2004does}, shared identity~\cite{ruilleryou}, assessing teleworkers’ performance~\cite{aboelmaged2012factors}, and impeded relationships with coworkers~\cite{gajendran2007good}. Among these challenges, decreased mental health and isolation is likely a key issue. 
In 2012, Bloom et al.~\cite{bloom2015does} conducted an experiment at the call center of CTrip, a NASDAQ-listed Chinese travel agency. According to the experiment, WFH promoted the performance of employees well due to the quiet environment, which reduced the attrition rate among home workers. However, 50\% of the participants decided to return to office after the experiment ended, because of concerns over being isolated at home. 
In addition, work-life balance~\cite{gold2013work} is also considered as another big challenge. Although WFH is often touted as a solution to work-life balance issues, conflict and tensions have been noted because it is hard to negotiate domestic activities and work-related activities as competing demands~\cite{adisa2017happened, maruyama2009multivariate}. For example, some people find that they have to work longer or more days in their work place at home because employers expect a higher productivity for those who WFH~\cite{maruyama2009multivariate}.


Although many of challenges may have been identified and described, we argue that the new context under both of the epidemic and current Chinese work culture could bring new challenges and reveal new insights. Firstly, WFH during an epidemic is at a sudden, national scale while in contrast, previous research focus more on WFH under normal circumstances when employees are allowed to WFH occasionally or irregularly~\cite{ruilleryou}. Before the epidemic there was no situation where 200 million people could be forced to suddenly WFH. The scale of this event could bring challenges for employees whose digital technologies and services in their companies could not support a large number of people WFH suddenly and simultaneously. Also, the epidemic could bring unexpected challenges in management regulations and systems (e.g. written report, online meetings, collaboration tools). For example, employees tended to be not familiar with WFH, their companies usually did not make WFH regulations and policies before the epidemic, they also experienced WFH with families especially parents during the epidemic. Consequently, WFH in China could be different and face more challenges than what has been previously documented. We have no answer about if a sudden and sustained requirement could promote employees to WFH after COVID-19 and what preparedness and actions we should make. In summary, our research is motivated by the need to understand  what challenges people could meet in this context. We documented and described what difficulties the Chinese encountered while WFH and then identified potential challenges and corresponding preparedness in design for both now and in the future.

\section{Methodology}
\subsection{Context, Research Questions and Participants}
We conducted semi-structured interviews with 12 participants with regard to people’s life experience from 24th of January 2020 to 30th of April 2020 in China. The research has been approved by the ethics review board in our research institution. Our goal is to understand \textbf{1) what were people’s WFH lives like during COVID-19 lockdown in China; 2) what kind of difficulties did Chinese people WFH encounter; 3) what were the main challenges and implications for WFH practice in the future.} We chose the start of this interview period in order to capture the experience of WFH during a lockdown where millions of people were affected while it was fresh in workers' memory \footnote{\url{https://www.worldometers.info/coronavirus/country/china/}}. The end date of the scope of the interviews was set on 30th of April 2020 because the increasing infected cases were under control generally and industries were resuming business gradually \footnote{\url{http://www.scio.gov.cn/zfbps/32832/Document/1681809/1681809.htm}}.

\begin{figure}[h]
  \centering
  \includegraphics[width=\linewidth]{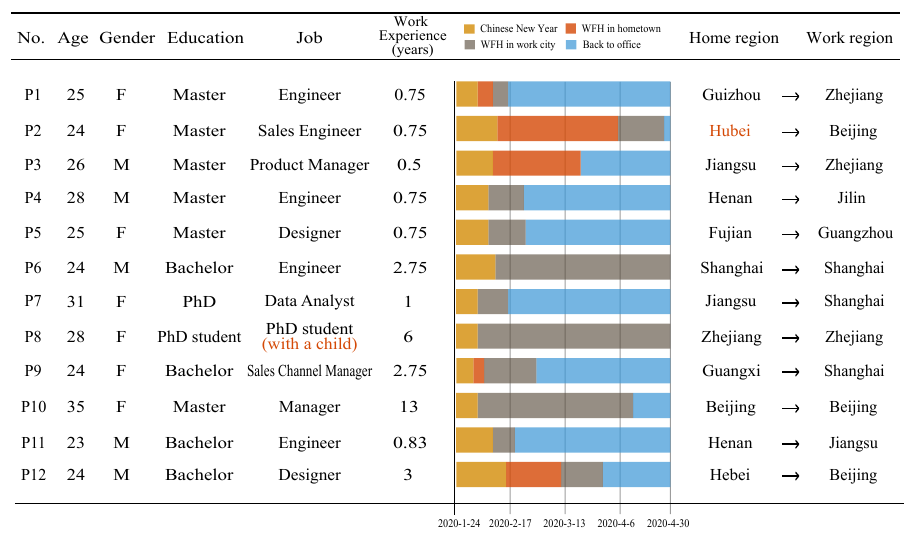}
  \caption{Basic information and traveling timeline of our participants' experience, from 24th of January 2020 until 30th of April 2020 (98 days). Work experience is indicated in years. P2 came from Hubei province, where COVID-19 was first reported in its capital city Wuhan, so stayed in her hometown for longer. P8's university was closed during our scoped time period so she was able to keep on working from home after the end of the lockdown. Both of P10's hometown and work city were in Beijing where the lockdown lasted longer.}
  \label{fig:traveltimetable}
\end{figure}

We recruited 12 participants (F=7, M=5) using the snowball sampling method, aged 23-35 (Mean = 26.4, Median = 25) and had work experience for 0.75-13 years (Mean=2.7, Median=0.9). Participants were initially recruited from the personal network of the first author, and then from recommendation from the early participants. They were conttacted on WeChat, and the interview was performed using WeChat call, as face-to-face communication was impossible due to a country-wide lockdown.

The recruitment criteria were: 1) a stable paid job (including PhD stipend) working in an office before 24th of January 2020; 2) WFH for at least 2 weeks between 24th of January and 30th of April 2020. Finally, we targeted mainly young professionals working in technology-centric companies, in line with previous work~\cite{belanger1999workers}, as the nature of their work makes it easier to work from home. Among our participants, 4 (P1, P2, P5, P11) were on a 996 schedule (or similar), meaning working from 9am to 9pm, 6 days per week.

Due to the practice of travelling to hometowns during Chinese New Year, we need to explain the timeline of participants' traveling timetable (See Figure~\ref{fig:traveltimetable}). Generally, workers plan to return to their work cities like Shanghai, Beijing and so on after the 7-day holiday. However, due to the lockdown, only P1, P9 went back to their work cities as usual successfully, and they needed to work from home in their work cities; P4, P5, P7, P9, P11 went back to their working cities after the extended Chinese New Year holiday and then started to WFH at working cities. P2, P3 stayed in their respective hometowns and practiced WFH for a longer time. P6, P8, P10 spent Chinese New Year in the same places as their working cities so they did not have to travel around different cities. Among the participants, the shortest period of WFH was 14 days and the longest was 91 days. 


Three participants (P1, P6, P9) were accustomed to online work before the pandemic because they worked with overseas subsidiaries of their company.
Our participants are mainly developers and designers (P1, P3, P4, P5, P6, P7, P11, P12) in IT companies, and the oldest participant (P10, 35 years old) was in management level. We also included two sales-related employees (P2, P9) and one PhD student (P8) who had a baby just before the epidemic.

\subsection{Data Collection and Analytic Approach}
All 12 interviews were semi-structured, conducted by the first author remotely via WeChat calls and recorded by phone and computer, after collecting interviewees consent. Interviews lasted from 40 minutes to 1.5 hours. Participants were paid 50RMB (around 7USD) by WeChat. The participants information is summarized in Figure~\ref{fig:traveltimetable}.

Based on Grounded Theory~\cite{charmaz2008constructionism}, three authors, all Chinese native speakers, used Microsoft Word to code the interview data iteratively, and then the coded data were discussed among all the team members to resolve conflicting interpretations. We derived our codebook after an initial coding phase done on 3 participants.

\section{Results}
\begin{figure}[t]
  \centering
  \includegraphics[width=0.8\textwidth]{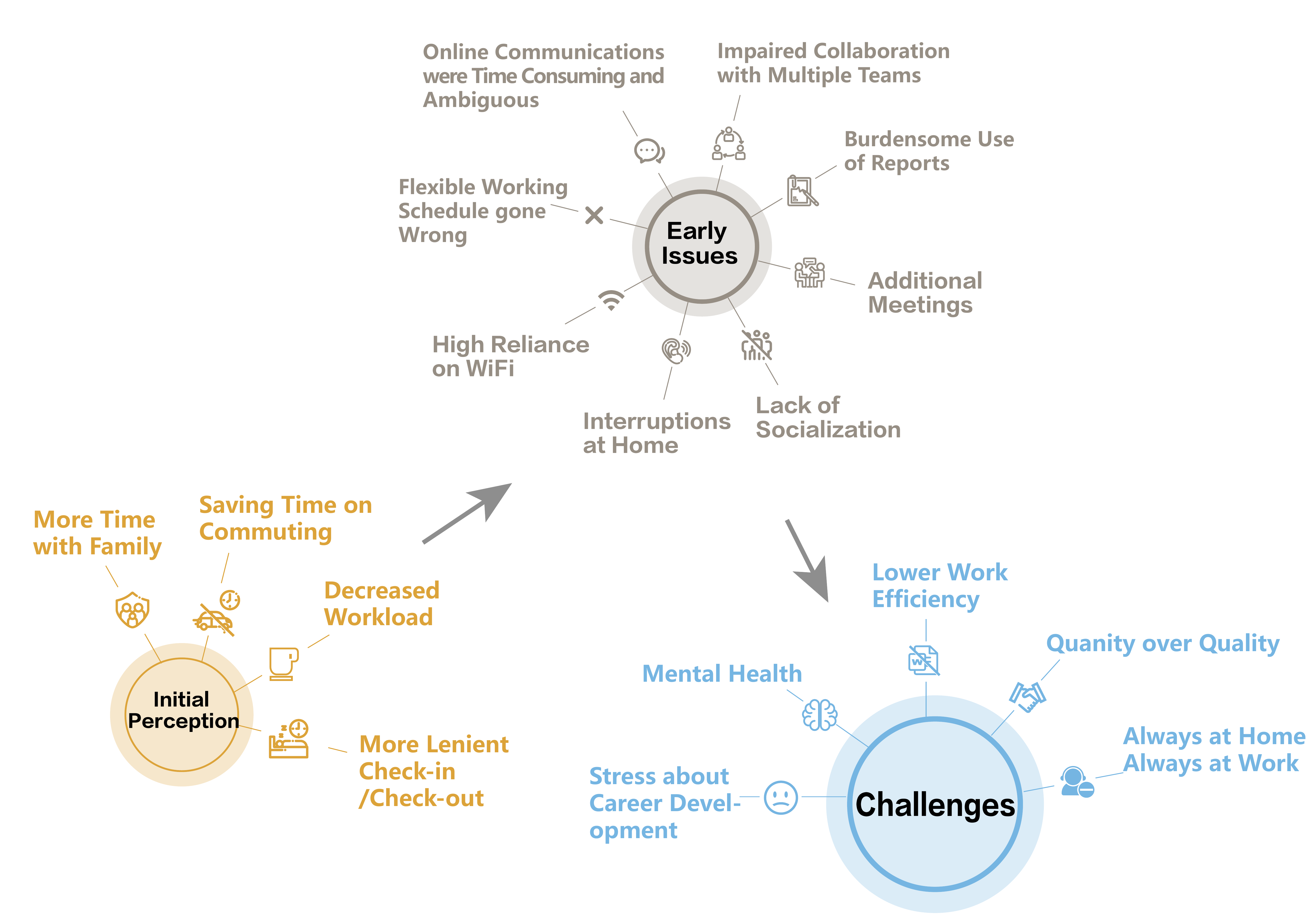}
  \caption{Summary of our results}
  \label{fig:results}
\end{figure}

We present the qualitative findings according to the themes we found in the data. The quotes have been translated from Chinese to English by the first author. Results are summarized in Figure~\ref{fig:results}.

\subsection{Background and Context}
For our participants, the change from working at the company’s office to WFH was sudden and happened during the Chinese New Year holiday, which is the most major holiday in China. At this time, most Chinese undertake annual travel to their respective home towns. The onset of the quarantine coincided with the start of these holidays, and therefore most of our participants were in their home towns, which were not the same as the cities of their work. Each province was given the choice of extending the public holiday for a longer duration depending on the lockdown needs of each province.
After the end of the official holiday, 4 of our participants remained in their home towns, living at their parents’ places to continue WFH. The remaining returned back to their work city at various points of time.

\subsection{Initial Perception of WFH}
Among the participants, the initial perception of WFH was unanimously positive: they reported that they suddenly had a reduced work load, and more time to spend with family, more flexibility and felt less closely monitored by their managers.

\subsubsection{Decreased Workload}

Half of the interviewees (P1,P2,P4,P7,P10,P12) reported that the activity of their company was negatively affected by the enforced quarantine. One cause was the timing that coincided with a normally quiet period at all work places. For many workers, this is the single vacation period of the year, meaning it is easy to slow down since most organizations are not working (P10). However, the second cause of decreased workload was the delay in resuming work and production when the quarantine was extended, which caused an overall slowdown in business (P1,P2). The final reason is an overall lack of preparedness for remote work at most companies, meaning that the first few days were spent trying to figure out how employees should be WFH: "I could not enter my "working state" in the first few days, and I ended up not working a lot." (P1).

\subsubsection{Saving Time on Commuting}

5/12 of the participants (P2,P3,P6,P10,P12) stated that another perceived benefit of WFH was a decrease in commuting time. When working at the office, P6 \& P10 reported a relatively long commuting time in one-way, between 40 and 120 minutes. P10 mentioned that she experienced a big difference between commuting time with emptier roads (40 mins) and with normally congested (2 hours). She declared “[...] I saved this time and could use it to exercise.” P12 said the saved commuting time could let him sleep until the last minute before clock-in. In contrast, P6 would use this time to catch up on social media, practice English and read the Economist on the subway. He had previously given up on renting a room close to the company because of the high cost. P6 was pleased with the WFH arrangements as “saving commuting time is a plus”. However, instead of undertaking non-work related activities, he used this saved time to start working earlier and to work longer in the evening. 

\subsubsection{Spending more Time with Family}

Working and living with their parents (4 participants) was initially seen as very positive. P1 described how her parents were curious about the content of work and her role at work: "The epidemic brought a chance for my family to know more about my work…It shows English on the screen, (so) they sit next to me to observe my screen quietly." (P1). P6 observed that his relationship with parents became better because he spent more time at home.

P3 reports a similar experience: “We ate together and watched TV every night during that period, it was nice". (P3). P8 also found it easier to take care of her baby “I need to feed my baby all the time. [Usually] it is not convenient to prepare milk for the baby because I need to freeze and store the milk but now [during epidemic] it is convenient” (P8). Her parents-in-law’s also moved to her house during the quarantine so she could get help and support while working on her PhD at the same time.

\subsubsection{More Lenient Check-in and Check-out Processes}
8 of our participants reported using a check-in and check-out system in the office before the quarantine. Usual check-in time would be between 9 and 10am, and checkout time would be between 6 and 9pm. Such long hours have a nickname in China, "996", which was controversially endorsed as tech culture by Jack Ma, founder of e-commerce giant, Ali Baba\footnote{https://edition.cnn.com/2019/04/15/business/jack-ma-996-china/index.html}. It means working from 9am to 9pm, 6 days a week. However, only 3 of the participants had to use a similar system during WFH, which in turn reduced the need to work at very specific hours. For them, the lack of a rigid schedule would give them more flexibility in planning their work day.

\subsubsection{Flexible Working Hours}

Four participants adapted their working hours while WFH. Some of them slept late and would thus start work later: "I couldn't get up in the morning so I would attend the meeting first and go back to sleep again. I did not start working until the afternoon." (P1). Some participants (7/12) kept to their usual schedule, in order to be able to collaborate with colleagues whenever called upon to.

For the reasons stated above, the initial WFH experience was mostly positive. 

\subsection{Early Issues}
While the first few days of WFH were rated rather positively by our interviewees, over time small issues started appearing. The very things seen as advantages turned out to have unexpected side effects.

\subsubsection{Flexible Working Schedule gone Wrong}
While we initially highlighted that participants were happy to have a chance to adapt their work schedule if needed, they also became dependent on other colleagues’ schedule. This evolved into being seen as being available anytime by colleagues, as long as colleagues are working. This was especially true in technology-centric companies, where working hours could be very long, and the average age of employees is younger\footnote{https://hub.packtpub.com/Chinese-tech-companies-dont-want-to-hire-employees-over-30-years-of-age/}. P2 explained that her “manager is 10 years older than me and goes to bed around 1:50am”, but since “most of my colleagues are young and not married, they tend to stay online even later”. This practice of working late echoes a similar phenomenon in Chinese offices pre-quarantine, where it is not seen as acceptable for employees to leave before their manager does. Additionally, employees want to avoid being to be the first one to leave work. However, during WFH, the participants experienced a drift in this timing, with nobody “leaving work” before their reporting officers, even if it was late in the night. 

\subsubsection{Online Communications were Time Consuming and Ambiguous}
Because of the lack of physical proximity, workers would use online communication for much of the day. P9 described his frustration that while collaborating with colleagues, "They may just put the phone there and you don't know whether your colleagues are listening or not." (P9). These audio calls would cost much time of team members to ascertain others’ level of assent or commitment to the topic. For example, "if someone forgot to enable audio, the leader would not know if [the worker] was in the meeting, thus [the leader] would ask 'Are you listening?'. In the office, [the leader] would just have a glance to understand others' states" (P3). 

Another issue arose from the switch from oral communications to written communications. In an office, a lot of communication is done orally. Instead, while WFH, participants reported using written communication more often, e.g. emails, instant messaging. The nature of these asynchronous communications could lead to misunderstandings, ambiguity, and in the long run, worsen the working atmosphere: “If employees did not understand the tasks [I give them] correctly the first time, they may not ask for clarification” (P10). Similarly, P11 reported that it is sometimes easy to explain something physically with words and gestures, but “It is not easy to explain [these] things online although we are using online collaboration tools.”

WFH makes the communication between different levels of the hierarchy longer. “My work needs to be reviewed by three different levels. If I submit a solution at around 8:00PM, my leader may give me a reply by 11:00PM. [After review by the senior leaders] I would get the final approval by 1:00AM.” When working from the office, communication could be sped up by meeting physically or physically sending the report to the supervisors.

Overall, while online communication still enables direct communication with colleagues and managers, it cannot replace face-to-face interaction and is usually slower (e-mail) or more ambiguous (voice call).

\subsubsection{Impaired Collaboration with Multiple Teams}

Online collaboration was not seen as efficient as offline communication, and the problem would only grow bigger when involving multiple colleagues and teams.
“Work communication efficiency is not as good as what you may imagine. My work is not independent and [requires close collaboration, thus] I need to coordinate with three or four parties... You have the downstream colleagues and upstream colleagues. The team needs to discuss and complete the work together. Communication is troublesome." (P7). 
The timetable of different team members were also tightly linked. If one link of the collaboration chain had a problem, the others links' work efficiency would be affected. P2 explains that this collaborative aspect decided her timetable. "We would get a [new] task in the evening. We would work until 12:00AM and even later. Not only we were online at that time, but our customers were also online."(P2). "You need to reply to other members [as quickly as possible] any time they ask you."(P5).

\subsubsection{Burdensome Use of Reports}

As the check-in and check-out procedures were relaxed, managers seemed to feel that they did not have much control on their subordinates. One way to mitigate that was to implement daily reports (4 participants) or weekly reports (5 participants). The goal was to be able to keep track of subordinates’ progress and be able to hold them accountable. P11 reported that “It would be hard to slack off if you put the minutes of your day precisely in the daily report.” Our participants did not welcome these additional reports as they felt these daily reports added extra work for them. The reporting duties were felt to be burdensome but unavoidable. P10 states: "They [employees] thought that submitting a work report is boring but (at least) the supervisor knows what employees have done… But if they did not write (the report), the communication process to overcome that would cost more time, which is also problematic." P10. 

\subsubsection{Additional Meetings}

In addition to regular work reports, some participants found that meetings became more frequent. Before the epidemic, 8 participants had at least weekly  meeting to discuss the work done and plan for the next iteration. Some participants reported the frequency of these meetings drastically increased after that. For example, P7 said she had five to six meetings every day during WFH instead of one meeting every day or every two days. But participants also saw the importance of such meetings, as these meetings “help employees report progress, [...] deal with problems, [...] adjust work arrangement, and improve communication within the team” (P1). P6 pointed out that it is a good way to “let everybody knows what others are doing”. What could and would usually be an informal discussion with the manager coming by the employee’s desk would turn into longer online meetings instead, and with the flexible working schedules, such meetings would also happen at night, outside of pre-quarantine working hours.

\subsubsection{High Reliance on WiFi}

5/12 of the Participants (P2,P3,P6,P11,P12) reported heavy reliance on their connection performance: limited connections would lead to problems for online communication, which would in turn reduce the amount of work done during the day. P3 could not use video communication much because the bandwidth at home was not sufficient to support it. P11 said: "The main problem is bandwidth; it requires high internet speed if the user needs to share screen or do video calls. The quality of video is poor". The poor network had a bad impact on work efficiency because sometimes they could hardly use online whiteboard or screen sharing which are very necessary tools in collaborative jobs such as software developers.
Poor internet connection, which happens more often in the smaller cities of their hometowns compared to larger ones, was another obstacle for the participants. Connectivity issues led to anxiety about not getting credit for their work, or being less visible to their manager as they could not use video.

\subsubsection{Interruptions at Home}

Initially, participants enjoyed spending time with their family, but WFH would also cause interruptions. It could be WiFi connection issues as seen above, or noise in the street, but also family interruptions. P11 stated that because of this experience, he would not want to work with family members around even after he gets married, because he could not focus his attention on the job and was disturbed frequently. P9, who lived with roommates, reported that “[as] we live in the same bedroom, we would disturb each other while taking voice calls”. P2 mentioned being interrupted by parents: “...your parents start talking with you [while working] about something, or they are talking, and you want to join them when you hear their discussion.”

Participants reported feeling always interrupted. There were multiple sources of interruptions: meetings with colleagues and managers, parents or children at home, or even getting sidetracked. "I started to work as soon as I got up, but I did not finish the work until the night. Because I might watch TV shows, or chatting while working." (P2), P11 expressed a similar experience: “Even though I do not have children and parents around me. I was (more) likely to stop working and do other things.” (P11).

\subsubsection{Lack of Socialization}

Because of the strictly enforced quarantine, our interviewees could either barely or not leave home at all. Thus, some participants’ mental health suffered from the lack of socialization: “The epidemic locked down my socialization.” (P10). P6 missed lunches with colleagues: "If I have lunch with my colleagues, we may not necessarily chat about work, but the contents of the conversation may also be very useful, which can expand my knowledge.” But despite the social isolation, participants also found other ways to socialize, such as playing online games with friend (P6, P11), or discussing with colleagues more often using WeChat (P4). For P6 specifically, it did not change the situation drastically, as “I do not socialize with others much even before the epidemic... As a ‘Homebody’, interacting with people online is enough for me.”

\subsection{Challenges}

\subsubsection{Quantity over Quality} 


With the shift from oral to written communication, workers became worried about their work output: managers would request written reports to assess the amount of work done. Participants found it hard to tell the truth on the work actually done. P11 said "What you wrote [for the report] depends on yourself". This could cause employees to over report the amount of work done and start focusing more on the quantity of work (i.e. work faster, and more superficially on many things) than quality. In the long run, this shift from quality to quantity put more pressure on employees and would push them to spend even more time on their reports. 

\subsubsection{Lower Work Efficiency}

Although it is hard to objectively compare work efficiency before and during WFH, 5 participants reported that their efficiency decreased while WFH. The initial relaxation of supervision processes also led some participants to sometimes work less: “My performance was different during different periods. When I did not want to work […], my efficiency was very low. But if I needed to rush something for a deadline, the stress pushed me to improve efficiency.” (P4). This lower perceived efficiency was also linked to the time lost on additional reports, meetings, and the overall lower efficiency of online communication. 

\subsubsection{Stress about Career Development}

Another big concern for our participants was that they became worried about their promotions in the future. P7 worried that the WFH episode had a bad influence on her career development plan. One key reason was that they did not have a good performance evaluation system when doing WFH. In addition, she feared that her manager would feel like she was not working hard enough: “If I could not find the leader/manager and submit my work, he would think I did not work” (P7). As the quarantine ended in various cities, P2 was still not back in the company’s city, and thus reported worrying about getting fired because of WFH: “Other colleagues have been back to work, but I am  still at home because I come from Hubei, which makes me anxious and worry about losing my job. Many companies have gone down and if my company wants to fire employees, I would likely be the first one.” (P2). Hubei was the province most affected by the virus containment measures, and had the strictest, longest lockdown period in China. 

\subsubsection{Always at Home, Always at Work}
Since there were no spatial boundaries between work and home, coupled with interruptions and other intrusion of privacy (e.g. staying on voice call with colleagues), participants complained about the “24 hours” schedule. In line with aforementioned “996” schedule in China, a well-known term for a “24 hours" schedule is “007”, which means work for 24 hours, 7 day per week. This is an obvious exaggeration but captures the perception of many of our participants.

P9 reported having trouble with her manager: "My leader thought I can be called at any time since I always stayed at home." (P9). P2 complained about receiving work assignments any time of the day: “Sometimes I don't have any task during the day, and then in the evening suddenly a task came, and I had to work until midnight to finish it." (P2).

\subsubsection{Mental Health}

The epidemic also affected mental health. From previous work~~\cite{cava2005experience}, we know that people placed in quarantine may show depressive symptoms. People in China were overall worried about the frightening evolution of the public health crisis. P4 was worried about getting infected: “when someone was suspected to be infected, they were brought to be quarantined immediately.” P1 reports that the epidemic was on “everyone’s mind”. P10, as a manager, also tried to give her subordinates some psychological reassurance.
P2 highlighted that “living with my parents when WFH brought [me] comfort but I was still sleepless at night and extremely depressed at the early stages of the epidemic”. P4 also reported increasing fear “while working in isolation, I think my fear was magnified 10 times.” (P4). P2 felt distressed with the daily bad news:  “Every day when I read the bad news, someone died, someone was infected. I felt very distressed.” (P2).

All of these issues combined together took their toll on our interviewees, and increased their levels of stress as compared to before the epidemic.



\section{Discussion}
We discuss the implications of our results for WFH, and how to potentially solve the identified challenges.

\subsection{Great Oaks from Little Acorns grow}
The WFH experience started positively for our interviewees, but small issues ended up raising important challenges. Ultimately, most of the challenges stemmed from earlier issues, and in turn, challenges contributed to the overall reported deterioration of the mental health of our participants. Some issues are also linked together and, as such, can hopefully be tackled together.
To make WFH manageable on a large scales, companies 
may need to address these problems holistically and in parallel, rather than focusing on just a few. 

\subsection{Documentation of Work Hours and Achievements}
Participants expressed a need for documentation of work hours and achievements when WFH.
Both employees and reporting officers would benefit from a system for documentation of work tailored for WFH. The benefit of such a system is that employees would be able to be accountable, to demonstrate efficacy and not jeopardize their promotions, while managers would be able to monitor their subordinates' work efficiently. However, there remain challenges regarding the details of the reporting. 
One challenge mentioned by our interviewees was that the existing mechanisms for accountability involved more meetings and reports, which would take time from their work time. While this points to a need to automatize reporting, there remains a gap between detectable data about documents used by employees, versus actual work achievements. Possible challenges would be defining what constitutes work achievements, when they should be reported and how they should be reported. Consequently, this necessitates a more organized approach to WFH than is currently implemented. 
The documentation of work achievements is complicated by the likelihood work hours creeping when WFH. Besides procrastination, employees are experiencing legitimate difficulties with focus when WFH, due to interruptions and the presence of other people in the home. Finally, in extreme contexts where WFH is necessitated by national emergencies, employees are additionally facing legitimate sources of stress such as food, shelter and work security.
Hence, more work is needed to explore the desirable features for a documentation of work system. Our results suggest that it might be important to allow the workers to have a control on what they share, so they that would focus more on quality than quantity. Another important aspect would be to make such a tool only for reporting what happens during negotiated work hours, which would avoid late night meetings and requirements from managers. Finally, some coordinated understanding is needed from all parties as to what might be a fair amount of work in times of national emergencies. 








    
\subsection{Comparison to Previous Work}
\subsubsection{Similarities}
Our results are in line with previous work, with similar challenges, such as mental health~\cite{ruilleryou}, reliance on technologies and the importance of good software tools~\cite{miele2020digital, raghuram2019virtual, ruilleryou}, as well challenges for managing workers~\cite{ruilleryou}. Nilles~\cite{nilles1997telework}  also suggested that time saved on commuting could be used for work instead, and was indeed the choice of one of our participant(P6).

\subsubsection{Differences}
Interestingly, Ruiller et al's~\cite{ruilleryou} results showed improved productivity from telework experience as well as a positive view on work hours flexibility. While we had similar results at the initial period of WFH, we observed that as the quarantine wore on, participants were reporting a lower work efficiency. For our participants, flexibility of working hours contributed to their feeling of always being at work while at home, and subsequently created a more negative sentiment about WFH. We believe that this difference is due to the contexts: in Ruiller et al's work, participants were only teleworking half to two days per week instead of full time. As such, they would still spend time at the office, and be able to reduce the impact of all the issues we identified such as 100\% online communication, additional reports and meetings. In contrast, our participants were spending almost 24 hours at home for 5.9 weeks in average (Median = 4.9 weeks). This sustained period suggests that without organizational and work platform adjustments, WFH is unsustainable and may be harmful to mental health in the long run. 

\subsubsection{Comparison with Bloom et al.}
Bloom et al.~\cite{bloom2015does}'s experiment with the Chinese Ctrip company in 2015 reported promising results on the experience of WFH. In their experiment, 249 call center workers worked from home for up to 9 months. They identified clear advantages (improved productivity, reduced commuting time, flexibility). Our study depicts a less positive image of WFH, but the difference can be explained mostly by the nature of participants' work: Bloom et al's~\cite{bloom2015does} used call center workers vs. our technology-centric company workers. Call center workers have strict shifts for work as opposed to our participants' usual varying days, with common overtime. The work performance metrics are also well defined in their case, e.g. duration spent on time, dropped calls proportion. 
Hence, our work suggests that the combination of sustained periods of remote work along with work that is not easily definable or quantifiable can lead to lower employee satisfaction in areas such as promotions and quality of life. 

\subsection{Limitations}
Our study has some limitations. First, we interviewed mostly people in China, who work in technology-centric companies, and who tend to be more educated. The abrupt transition from office to home was thus likely easier for that specific population than for less tech savvy users. Our interviewees were also overall young, early career workers, and thus more open to changes. Similarly, the type of work they did allowed them to still perform most of their tasks from home, which would have been more problematic in other lines of work.

\subsection{Effects of the COVID-19 Epidemic}
The COVID-19 epidemic impacted our interview results. The most obvious effect is the effect of the epidemic on mental health. In addition, most of our interviewees reported a decreased workload because of the national holiday followed by extended lockdown, and reduced socialization due to the ongoing lockdown.
Despite these effects, we believe that the noise induced by the epidemic is outweighed by the unique opportunity of investigating a major, sudden, large scale WFH experience, where millions of workers started WFH overnight for several weeks at once.
We believe that the timing and location of this study offers the unique opportunity to get a clear picture on how the transition between WFH and working from the office would happen, and of the effect of sustained WFH at scale.  

\section{Implications for Design}

\subsection{Negotiated Organizational Targets between Employees and their Supervisors}
From our results, it is clear that current management systems are not adequate for WFH. Existing monitoring systems are not ideal for monitoring work achievements for people whose achievements are not easily quantified. From our discussion, what is clear is that systems that represent a single view (e.g. just the organization's) will not be viable as it seems to require negotiation on several points to agree upon what is work. An example of the limitations of a top-down approach exist. In 2018, DingTalk for office promised to "boost productivity through better monitoring of employee movements and faster responses to important messages". However, the reception to DingTalk even during non-quarantine times was negative, with office workers complaining about how it "fosters an unhealthy work culture", is "inhumane and fosters distrust". Features that fuelled this discontent were endless notifications, its use for monitoring workers, and required reporting\footnote{\url{https://www.reuters.com/article/us-china-alibaba-labour/ding-always-on-alibaba-office-app-fuels-backlash-among-chinese-workers-idUSKBN1KO0RR}}. In our results, we observed a similar resistance. We conclude that in order for work platforms to meet the needs of organizations during extended periods of WFH, it is likely important to support negotiated work hours and work targets, and to recognize impaired efficacy when an emergency is the cause of WFH.  

\subsection{Preparing a WFH Transition}
For companies who want to transition to WFH, or in case of another epidemic where WFH becomes mandatory, we would advocate for companies to consider new management methods and tools. Managers and workers would thus have time to get familiar with the new procedures while everybody is still on site, thereby addressing the working hours creep and online communication issues. This might give enough time to managers and workers to get used to these new tools and use them in a healthy way (i.e. not outside work hours, etc...). This should in turn ease the quantity over quality challenges, and reduce stress over career development, which should be beneficial for workers' mental health. In essence, we recommend that when companies implement WFH training or rehearsal days, it should involve more than learning how to use remote communication platforms. Our study shows that WFH preparations should be built around organizational changes at all levels of the hierarchy. Perhaps most absent from current WFH protocols is explicit guidelines dealing with the new stressors that often surround the need to WFH. This study is sited in the context of an epidemic, but future drivers of WFH could include climate emergencies or security challenges. 

\subsection{Importance of Context}
Implications for designing systems that support WFH need to be aware of existing sociocultural practices of work. In our case, the outcomes were greatly influenced by the following elements:

\subsubsection{Type of Work}
As discussed, some types of work (e.g. call-center) can easily be done from home. Bloom et al.~\cite{bloom2015does} showed that WFH can be an overall positive experience for call-center workers in China. Our results suggest that WFH can have opposite outcomes on perceived work efficiency, despite the same geographical context. This can be explained by the nature of the work, the lack of quantifiable performance metrics for the job, leading to more but not necessarily effective monitoring.

\subsubsection{Work Culture}
Another element is the prevailing work culture. In our study, we studied workers from technology-centric companies, infamous for their 996 work pattern of high number of hours of work. While the term "996" is uniquely Chinese, a International Labour organization report from 2018 shows that the Arab States, East Asia and South Asia have the longest working hours, with more than 43\% of workers working longer than 48 hours a week \cite{griffin_ensuring_2018}. 


Hence, while the impact of work culture may be hard to compare across countries, and would need more cross-culture studies, we suggest that future work remain sensitive to the current practice of work and employment while doing research on WFH.

\subsubsection{Life Outside of Work}
Four of our participants stayed with their parents while WFH. It is a common situation in China, especially in large cities, where young professionals prefer to stay with their parents until they get married, 
for monetary and cultural reasons. Multi-generational homes like these are prevalent in many parts of Asia (e.g. Thailand, Singapore, Malaysia), for similar reasons~\cite{liu_families_2014}.
In contrast, more of their Western counterparts live in their own apartments, either alone or with their partners. This reduced the likelihood of interruptions experienced at home, which turned out to be a significant problem for our interviewees.

\section{Conclusion}
In this paper, we identified the challenges encountered by Chinese young professionals during the initial COVID-19 epidemic in China, where workers had to WFH on a large scale with little to no preparation. We conducted semi-structured interviews, and found out that while the initial WFH experience was positive, small issues arose and became major challenges over a short period of time. Our results bears similarities with other studies on remote work, but also pointed out differences. Our participants had less perceived efficiency, and the increase in flexible work hours, usually seen as an advantage in previous work in China~\cite{ruilleryou} evolved into larger concerns for our participants. We finally provide guidelines and recommendation for future work.

\bibliographystyle{ACM-Reference-Format}
\bibliography{sample-base}


\appendix
\end{document}